\documentclass{aa}  

\usepackage{natbib}
\bibpunct{(}{)}{;}{a}{}{,} 
\usepackage{graphics}   
\usepackage{graphicx}   
\usepackage{subfig}
\usepackage{epstopdf}
\usepackage{amsmath} 
\usepackage[varg]{txfonts}
\usepackage{pdflscape}
\usepackage{hyperref}   
\usepackage{longtable}
\usepackage{balance}
\usepackage{color}
\usepackage[normalem]{ulem}

\hypersetup{
    bookmarks=true,         
    colorlinks=true,       
    linkcolor=blue,          
    citecolor=blue,        
    filecolor=blue,      
    urlcolor=blue           
}

\def\ga{\,\,\raise0.14em\hbox{$>$}\kern-0.76em\lower0.28em\hbox{$\sim$}\,\,}
\def\la{\,\,\raise0.14em\hbox{$<$}\kern-0.76em\lower0.28em\hbox{$\sim$}\,\,}

\def\Msun{$M_{\odot}$}

\def\cm3{cm$^{-3}$}

\def\chem#1#2{$\mathrm{^{#2}\kern-0.8pt#1}$}
\def\reac#1#2#3#4#5#6{$\mathrm{\, ^{#2}\kern-0.8pt{#1}\, ({#3}\, ,{#4})\, {}^{#6}\kern-0.8pt{#5}\, }$}

\def\nuc{\mathrm{nuc}}

\def\be{\begin{equation}} 
\def\ee{\end{equation}}
\def\beqy{\begin{eqnarray}}
\def\eeqy{\end{eqnarray}}
\def\bmlet{\begin{mathletters}}
\def\emlet{\end{mathletters}}

\begin{document}

\title{The intermediate neutron capture process}
\subtitle{VI. Proton ingestion and i-process in rotating magnetic asymptotic giant branch stars }

\author{A. Choplin\inst{1,2}   
\and 
L. Siess\inst{1,2}
\and
S. Goriely\inst{1,2}
\and
P. Eggenberger\inst{3}
\and
F. D. Moyano\inst{4}
}
\offprints{arthur.choplin@ulb.be}

\institute{
Institut d'Astronomie et d'Astrophysique, Universit\'e Libre de Bruxelles,  CP 226, B-1050 Brussels, Belgium
\and
BLU-ULB, Brussels Laboratory of the Universe, blu.ulb.be
\and
D\'epartement d’Astronomie, Universit\'e de Genève, Chemin Pegasi 51, 1290 Versoix, Switzerland
\and
Yunnan Observatories, Chinese Academy of Sciences, Kunming 650216, China
}

\date{Received --; accepted --}

\abstract
{
The intermediate neutron-capture process (i-process) can occur during proton ingestion events (PIEs), which may take place in the early evolutionary phases of asymptotic giant branch (AGB) stars.
}
{
We investigate the impact of rotational and magnetic mixing on i-process nucleosynthesis in low-metallicity, low-mass AGB stars.
}
{
We computed AGB models with [Fe/H] = $-2.5$ and $-1.7$ and initial masses of 1 and 1.5~\Msun\ using the \textsf{STAREVOL} code, including a network of 1160 nuclei coupled to transport equations. 
Rotating models incorporate a calibrated Tayler-Spruit (TS) dynamo to account for core rotation rates inferred from asteroseismic observations of solar-metallicity sub-giants and giants. 
Initial rotation velocities of 0, 30, and 90~km~s$^{-1}$ were considered, along with varying assumptions for magnetic mixing.
}
{
Rotation without magnetic fields strongly suppresses the i-process due to the production of primary $^{14}$N, which is subsequently converted into $^{22}$Ne -- a potent neutron poison during the PIE. Including magnetic fields via the TS dynamo restores the models close to their non-rotating counterparts: strong core-envelope coupling suppresses shear mixing and prevents primary $^{14}$N synthesis, yielding i-process nucleosynthesis similar to non-rotating models. We also find that rotational mixing during the AGB phase is insufficient to affect the occurrence of PIEs.
}
{
Proton ingestion event-driven nucleosynthesis proceeds similarly in asteroseismic-calibrated magnetic rotating AGB stars and non-rotating stars, producing identical abundance patterns.
}


\titlerunning{}

\authorrunning{A. Choplin et al. }

\maketitle

\section{Introduction}
\label{sect:intro}

Understanding the chemical evolution of the Universe requires identifying the nucleosynthetic processes and their astrophysical sites. Elements heavier than iron are mainly produced through the slow (s) and rapid (r) neutron-capture processes, which occur at neutron densities of $N_n \sim 10^5$–$10^{10}$ and $N_n > 10^{20}$ cm$^{-3}$, respectively \citep[e.g.][]{arnould20}. The s-process primarily takes place in asymptotic giant branch (AGB) stars and in the helium-burning cores of massive stars \citep[e.g.][and references therein]{lugaro23}, while the r-process is associated with more extreme environments such as neutron star mergers, collapsars, or magnetorotational supernovae \citep[e.g.][]{arnould07,wanajo14,siegel19}.

In addition to the s- and r-processes, an intermediate neutron-capture process (i-process) has been proposed \citep[][]{cowan77}, operating at neutron densities of $N_n \sim 10^{13}$–$10^{16}$ cm$^{-3}$. Its existence is supported by stars showing chemical abundance patterns inconsistent with either the s- or r-process alone, but well reproduced by i-process models (r/s stars; e.g. \citealt{mishenina15, roederer16, karinkuzhi21, hansen23}). Evidence for proton ingestion and i-process nucleosynthesis may also be present in the Sakurai object \citep{asplund99,herwig11} and in certain pre-solar grains \citep{fujiya13,liu14,choplin24let}.

\begin{figure*}[t]
\centering
\includegraphics[width=1.8\columnwidth]{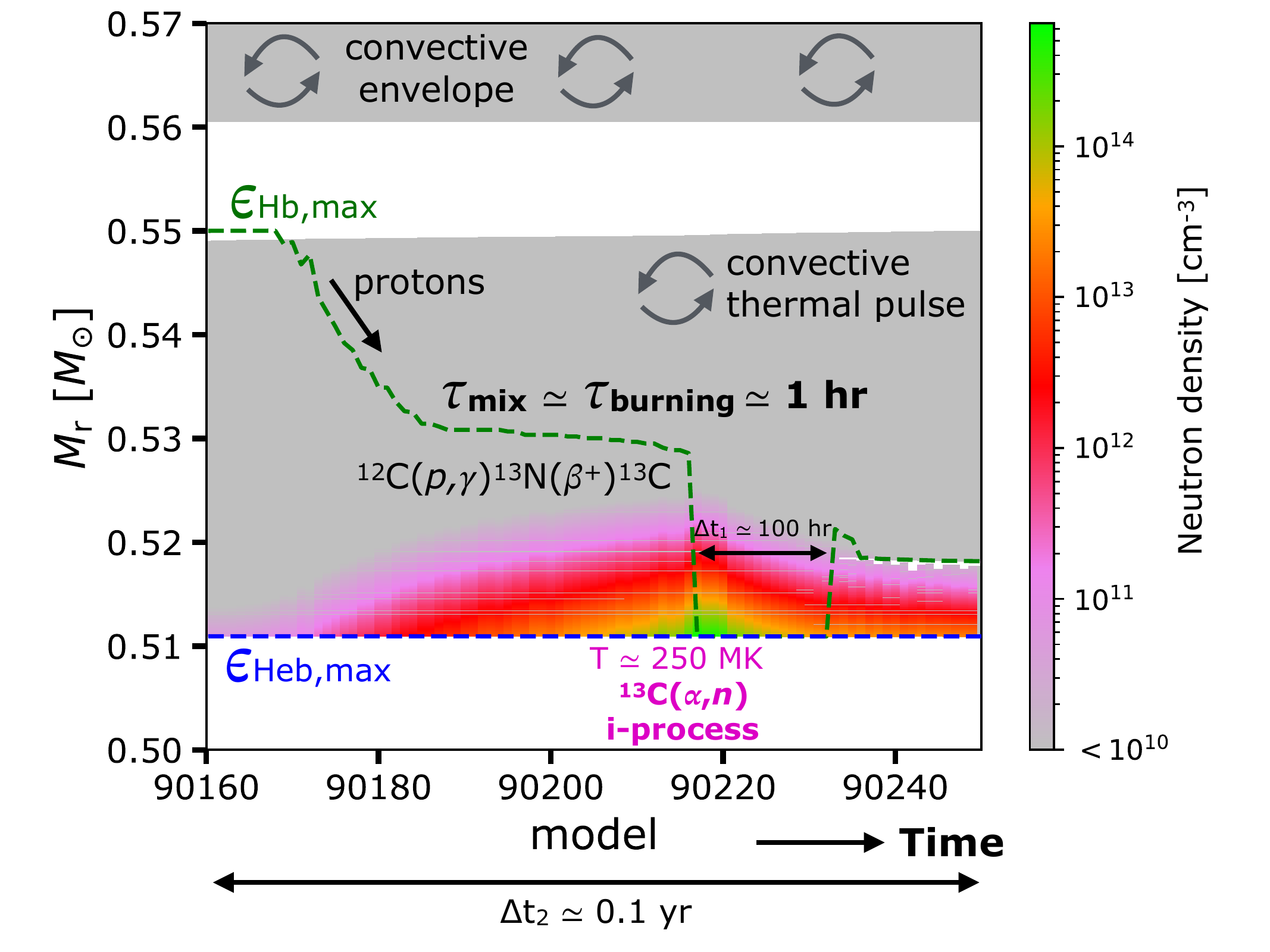}
\caption{ 
Kippenhahn diagram illustrating key features of a PIE in a non-rotating 1 \Msun, [Fe/H]~$=-2.5$ AGB model, computed with the STAREVOL code. Grey regions indicate convective zones. The dashed green (blue) line marks the location of maximum H-burning (He-burning) energy. The colour map shows the neutron density. The time between the peak neutron density and the split is $\Delta t_1 \simeq 100$~hr, while the total duration of the sequence shown is $\Delta t_2 \simeq 0.1$ yr.
}
\label{fig:pie}
\end{figure*}

The i-process can be triggered when protons are ingested into a convective helium-burning zone. This is expected in several astrophysical sites \citep[see][for a detailed list]{choplin21}, notably low-metallicity, low-mass AGB stars \citep[e.g.][]{iwamoto04, cristallo09a, suda10, ritter18b, choplin22a}. 
The proton ingestion event (PIE) mechanism is illustrated in Fig.\ref{fig:pie}. During a PIE, at the beginning of the thermally pulsing AGB phase, protons are mixed into the convective thermal pulse, captured through $^{12}$C($p,\gamma$)$^{13}$N, and rapidly converted to $^{13}$C via $\beta$-decay. The subsequent $^{13}$C($\alpha,n$)$^{16}$O reaction at $\sim 250$ MK produces neutron densities up to $N_n \sim 10^{15}$ cm$^{-3}$, triggering an i-process nucleosynthesis and potentially leading to actinide production \citep{choplin22b}. After the neutron density peak (green zone in Fig.\ref{fig:pie}), the i-process material is rapidly mixed throughout the entire thermal pulse, before it splits (at $M_{\rm r} \simeq 0.52$ and model $\simeq 90235$ in Fig.~\ref{fig:pie}). The processed material is ultimately dredged into the envelope and expelled through stellar winds.

Most AGB star models do not include rotation, although such stars as AGBs do rotate \citep[e.g.][]{vlemmings18}. 
Rotation affects stellar evolution by distorting the stellar surface at high speeds and by driving instabilities that transport angular momentum (AM) and chemical elements \citep[e.g.][]{heger00, maeder12}. Early studies suggested that rotational mixing could create a $^{13}$C-pocket \citep{langer99}. However, detailed models showed that the pocket was too small to reproduce observed s-process enrichments and that extra $^{14}$N synthesized thanks to rotation acts as a neutron poison, quenching s-process efficiency \citep{herwig03b, siess03, siess04}. Later, \cite{piersanti13} computed yields for rotating AGB stars and confirmed that rotation leads to the contamination of the $^{13}$C-pocket by the poisonous $^{14}$N. However, varying the initial rotation rates ($10 - 120$~km s$^{-1}$) and mixing efficiencies, they showed that a wide range of abundance patterns can be obtained.

Asteroseismic studies of sub-giant and red giant stars have revealed their internal and envelope rotation rates \cite[e.g.][]{beck12, deheuvels12, deheuvels20, mosser12, dimauro16, triana17, gehan18, tayar19,gangli24,mosser24,dhanpal25}. 
Comparisons with rotating models that include only AM transport by hydrodynamic instabilities provide too little coupling between the core and the envelope to reproduce the observed core rotation rates \citep{eggenberger12b, marques13, ceilier13, tayar13, cantiello14, eggenberger17, eggenberger19, moyano22}. This demonstrates that at least one additional, efficient AM transport mechanism is missing in the radiative zones of red giants. 
To match observed core rotation rates, \citet{denhartogh19} introduced an artificial, constant viscosity term in their solar metallicity, 2~\Msun\ rotating AGB models. They found that s-process production remained similar to non-rotating cases due to the strong reduction of rotationally induced mixing.

Motivated by the work of \cite{fuller19}, \cite{eggenberger22} proposed a modified version of the Tayler-Spruit (TS) dynamo \citep{spruit02} able to reproduce asteroseismic observations of sub-giant and red giant stars. 
This calibrated version better matches the core and surface rotation of low-mass ($1.4 - 2$ \Msun) Gamma Doradus pulsators \citep{moyano23a} than models without magnetic fields \citep{ouazzani19}. Surface boron abundances in massive stars are also better accounted for when including the modified TS dynamo \citep{asatiani25}.

This study is the first to investigate the impact of rotation on i-process nucleosynthesis in AGB stars. To account for the asteroseismic constraints available for red giant stars, we implemented the modified TS dynamo proposed by \cite{eggenberger22}. Sect.~\ref{sect:inputs} describes the models and physical ingredients, while Sect.~\ref{ref:res_calib} discusses the calibration of the dynamo. Results are presented in Sects.~\ref{sect:rotevol} and \ref{sect:ipro}, and conclusions are summarized in Sect.~\ref{sect:concl}.

\begin{table*}[t]
\scriptsize{
\caption{Main characteristics of the models computed in this work.  
\label{table:1}
}
\begin{center}
\resizebox{14.5cm}{!} {
\begin{tabular}{lccccccc} 
\hline
Model label & $M_{\rm ini}$  & [Fe/H] &$Z$   & $v_{\rm ini}$ & Magnetic & $C_{\rm T}$   & $N_{\rm n, max}$  \\
   &    [\Msun]        &        &   & [km~s$^{-1}$] & & &  [cm$^{-3}$]  \\
\hline
\texttt{M1.0z2.5\_v00}   &   1.0  & $-2.5$     &  $4.3\times 10^{-5}$       &   0    & NO & 0 & 3.0e14 \\
\texttt{M1.0z2.5\_v30}   &  1.0   & $-2.5$    &  $4.3\times 10^{-5}$     &  30     & NO & 0  &  8.1e13   \\
\texttt{M1.0z2.5\_v30\_CT1}   &  1.0   & $-2.5$    &  $4.3\times 10^{-5}$     &  30 & YES  & 1  &  5.5e14  \\
\texttt{M1.0z2.5\_v30\_CT50}   &  1.0   & $-2.5$    &  $4.3\times 10^{-5}$     &  30  & YES & 50   & 8.5e14 \\
\texttt{M1.0z2.5\_v30\_CT216}   &  1.0   & $-2.5$    &  $4.3\times 10^{-5}$     &  30  & YES & 216   & 3.6e14 \\
\texttt{M1.0z2.5\_v90\_CT50}   &  1.0   & $-2.5$    &  $4.3\times 10^{-5}$     &  90  & YES & 50   & 2.2e14 \\
\texttt{M1.0z2.5\_v90\_CT216}   &  1.0   & $-2.5$    &  $4.3\times 10^{-5}$     &  90  & YES & 216   & 3.4e14 \\
\hline 
\texttt{M1.0z1.7\_v00}   &  1.0   & $-1.7$    &  $2.7\times 10^{-4}$     &  0   & NO & 0    &  2.9e14 \\
\texttt{M1.0z1.7\_v30}   &  1.0   & $-1.7$    &  $2.7\times 10^{-4}$     &  30   & NO & 0    &  2.1e14 \\
\texttt{M1.0z1.7\_v90\_CT50}   &  1.0   & $-1.7$    &  $2.7\times 10^{-4}$     &  90   & YES & 50   & 1.7e14 \\
\texttt{M1.0z1.7\_v90\_CT216}   &  1.0   & $-1.7$    &  $2.7\times 10^{-4}$     &  90   & YES & 216   & 2.5e14 \\
\texttt{M1.5z1.7\_v00}   &  1.5   & $-1.7$    &  $2.7\times 10^{-4}$     &  0    & NO & 0 &  3.3e14 \\
\texttt{M1.5z1.7\_v90\_CT50}   &  1.5   & $-1.7$    &  $2.7\times 10^{-4}$     &  90   & YES & 50  & 6.2e14  \\
\texttt{M1.5z1.7\_v90\_CT216}   &  1.5   & $-1.7$    &  $2.7\times 10^{-4}$     &  90   & YES & 216  & 7.2e14 \\
\hline
\end{tabular}
}
\tablefoot{
Indicated are the model label, initial mass ($M_{\rm ini}$), [Fe/H] ratio, initial metallicity by mass fraction ($Z$), initial velocity at the zero-age main sequence, presence of magnetic fields (via the generalized TS dynamo; see Sect.~\ref{sect:transpang}), the $C_{\rm T}$ constant, and the maximum neutron density reached during the PIE.
}
\end{center}
}
\end{table*}

\begin{figure}[t]
\centering
\includegraphics[width=1\columnwidth]{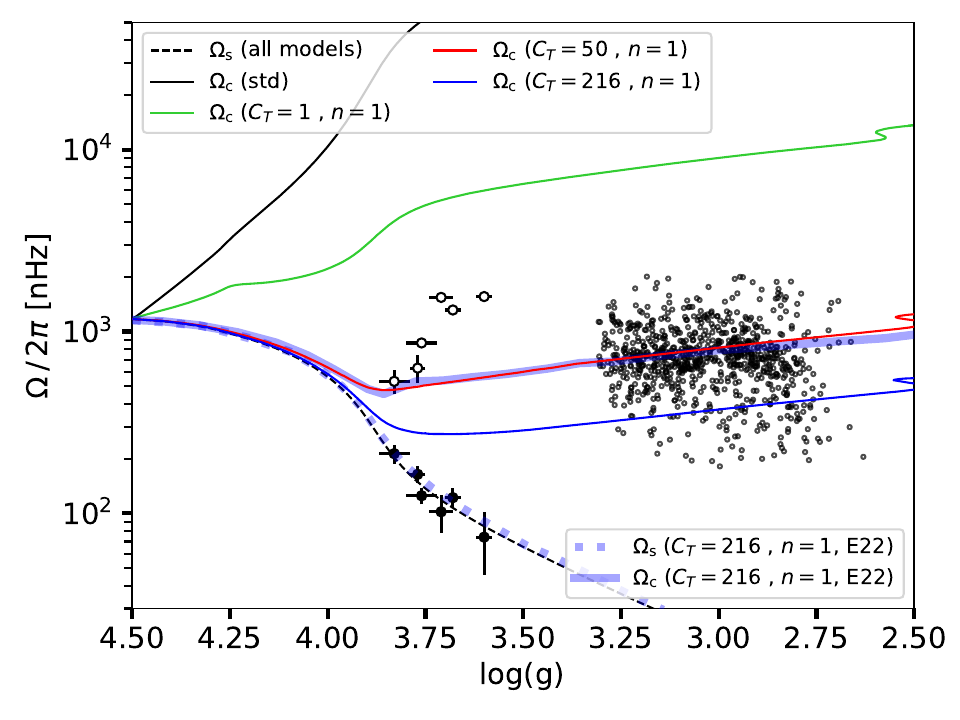}
\caption{Surface (dashed lines, $\Omega_{\rm s}$) and core (solid lines, $\Omega_{\rm c}$) rotation rates as a function of surface gravity for 1.1~\Msun, solar-metallicity models with $v_{\rm ini} = 5$~km~s$^{-1}$. Models are computed without (black) and with the Tayler instability, using $n=1$ and various values of the calibration constant $C_{\rm T}$ (green, red, and blue). The thick blue lines show the surface and core rotation of the 1.1~\Msun, solar-metallicity model from \cite{eggenberger22} (E22) computed with GENEC ($n=1$, $C_{\rm T}=216$). Large black-filled (open) symbols indicate observed surface (core) rotation rates of sub-giant stars \citep{deheuvels14}, while the smaller black dots show core rotation rates of RGB stars \citep{gehan18}.
}
\label{fig:omega_logg}
\end{figure}

\section{Physical inputs of the models}
\label{sect:inputs}

The stellar models were computed with the STAREVOL code \citep[][and references therein]{siess00, siess06, goriely18c}. As a first step, 1.1~\Msun\ solar-metallicity models were computed to calibrate rotational mixing efficiency (Sect.~\ref{ref:res_calib}). Then, 1 and 1.5~\Msun\ models at [Fe/H]~$=-2.5$ and $-1.7$, with $0 < v_{\rm ini} < 90$~km~s$^{-1}$ and varying strengths of magnetic mixing (controlled by the constant $C_{\rm T}$; see Eq.~\eqref{eq:dmag} in Sect.~\ref{sect:transpchem} and Table~\ref{table:1}) were computed. The solar(-scaled) composition was taken from \cite{asplund09}. Mass loss was treated using the prescription from \cite{schroder07} up to the AGB phase and from \cite{vassiliadis93} thereafter. As in previous studies, opacity variations due to molecular formation were included once the star became carbon-rich \citep{marigo02}. A mixing-length parameter of 1.75 was adopted. Nuclear burning was followed with a 411-isotope network (from $^{1}$H to $^{211}$Po) up to the onset of a PIE, at which point we switched to an i-process network of 1160 isotopes (from $^{1}$H to $^{253}$Cf) coupled with transport equations. Convective overshooting was included from the AGB phase onwards, at the top of thermal pulses (Sect.~\ref{sect:os}). 
For additional details on the physical inputs, especially the nuclear reaction networks and reaction rates, we refer to \cite{choplin21, choplin22a} and \cite{goriely21}.

Proton ingestion events generally occur during the first thermal pulses, dredging metals to the AGB surface and thereby increasing opacity and mass loss \citep[e.g. Sect.~3.3 in][]{choplin22a}. For the low initial masses considered here ($1–1.5$~\Msun), this results in rapid envelope loss, premature termination of the thermal pulse AGB phase, and suppression of s-process nucleosynthesis, which is thus not addressed in this study.

\subsection{Transport of angular momentum}
\label{sect:transpang}

Angular momentum transport is solved simultaneously with stellar structure equations. The equation for AM transport is purely diffusive and can be written as
\begin{equation}
\frac{\partial r^{2}\Omega}{\partial t}  =  \frac{\partial}{\partial m_r} \left[\left(4\pi r^{2} \rho \right)^{2} D_{\rm ang} \, r^{2}\frac{\partial\Omega}{\partial m}\right],
\end{equation}
with
\begin{equation}
D_{\rm ang} = D_{\rm conv} + D_{\rm over} + D_{\rm shear} + D_{\rm mag}
\label{eq:dang},
\end{equation}
where $D_{\rm conv}$ is the diffusion coefficient due to convection (derived from the mixing-length theory), $D_{\rm over}$ the overshoot coefficient (see Sect.~\ref{sect:os}), and $D_{\rm shear}$ the secular shear coefficient from \cite{maeder97}, which can be expressed as \citep{meynet13}
 \begin{equation}
D_{\rm shear} =  \frac{H_p}{g\delta} \frac{ K }{(\nabla_{\rm ad} - \nabla) + \frac{\varphi}{\delta}\nabla_\mu } \left(\frac{9 \pi}{32} \Omega \frac{d \ln \Omega}{d \ln r}\right)^2
\label{eq:dshm97},
\end{equation}
with $H_p$ being the pressure scale height, $K$ the thermal diffusivity, $\delta = -\left(\frac{\partial \ln \rho}{\partial \ln T}\right)_{P,\mu}$, $\varphi = \left(\frac{\partial \ln \rho}{\partial \ln \mu}\right)_{P,T}$, $\nabla$, $\nabla_{\rm ad}$, and $\nabla_{\mu}$ the temperature, adiabatic, and mean molecular weight gradients, respectively.
The $D_{\rm mag}$ coefficient in Eq.~\ref{eq:dang} denotes the viscosity associated with AM transport by the Tayler instability described in Sect.~\ref{sect:tsdyn}.
  
\subsection{Transport of chemicals}
\label{sect:transpchem}

The abundance $X_i$ of a nucleus $i$ is followed by solving the diffusive and nuclear burning reaction equation
\begin{equation}
\frac{\partial X_i}{\partial t}= \frac{\partial}{\partial m_r}  \left[\,(4 \pi r^2 \rho)^2\,D_{\rm chem}\ \frac{\partial X_i}{\partial m_r}\right] + \frac{\partial X_i}{\partial t}\bigg|_\nuc \ 
\label{eq_dif},
\end{equation}
where 
\begin{equation}
D_{\rm chem} = D_{\rm conv} + D_{\rm over} + D_{\rm shear} + D_{\rm eff}
\label{eq:dchem},
\end{equation}
with $D_{\rm conv}$, $D_{\rm over}$, and $D_{\rm shear}$ as in Sect.~\ref{sect:transpang} and $D_{\rm eff}$ the effective mixing coefficient from \cite{chaboyer92},
\begin{equation}
D_{\rm eff} = \frac{1}{30}\frac{|rU(r)|^2}{D_h},
\label{eq:Deff}
\end{equation}
where $U(r)$ is the amplitude of the radial component of the meridional velocity \citep{maeder98} and $D_{\rm h}$ the horizontal (i.e. on an isobaric surface) shear diffusion coefficient from \cite{zahn92}. The $D_h$ coefficient can be expressed as
\begin{equation}
D_h = \frac{1}{c_h} r \left| 2V - \alpha U(r) \right|,
\end{equation}
with $\alpha = \frac{1}{2}\frac{\text{d}\ln(r^2\bar{\Omega})}{\text{d}\ln r}$ and $c_{\rm h} = 1$, $\bar{\Omega}$ being the average value of $\Omega$ on an isobar and $V$ the horizontal component of the meridional circulation. 

The first and second terms on the right-hand side of Eq.~\ref{eq_dif} account for the changes resulting from diffusive transport and nuclear burning, respectively.
During a PIE, the convective and nuclear burning timescales become similar (about 1~hr), requiring nucleosynthesis and transport of chemical species to be solved simultaneously \citep[cf. Sect.~2.1 in][]{choplin22a}. 
Unlike AM, no magnetic term is included for the transport of chemicals (Eq.~\ref{eq:dchem}) because the chemical mixing caused by the Tayler instability is likely negligible \citep{fuller19}.

 \subsection{Asteroseismic-calibrated version of the Tayler-Spruit dynamo }
\label{sect:tsdyn}

Following \cite{eggenberger22}, the $D_{\rm mag}$ coefficient in Eq.~\ref{eq:dang} can be written as 
\begin{equation}
D_{\rm mag} = \frac{\Omega \, r^2}{q} \, \left( C_T \, q \, \frac{\Omega}{N_{\rm eff}} \right)^{3/n} \, \left( \frac{\Omega}{N_{\rm eff}} \right)
\label{eq:dmag},
\end{equation}
with $q = \left| \frac{\partial \ln \Omega}{\partial \ln r} \right|$ the shear parameter, $N_{\rm eff}$ the effective Brunt-V{\"a}is{\"a}l{\"a} frequency given by
\begin{equation}
N_{\rm eff}^2 = \frac{\eta}{K}N_{T}^2 + N_{\mu}^2, 
\end{equation}
where $N_T$ and $N_{\mu}$ denote the thermal and chemical composition
components of the Brunt–Väisälä frequency, and $\eta$
the magnetic diffusivity. 
The $C_T$ quantity is a dimensionless calibration parameter accounting for uncertainties in the damping timescale of the azimuthal magnetic field. The $n$ parameter distinguishes between different prescriptions: the original TS dynamo corresponds to $n=1$, $C_T=1$, while $n=3$, $C_T=1$ recovers the prescription of \cite{fuller19}. The asteroseismic calibration of \cite{eggenberger22} instead adopts $n=1$ and $C_T=216$ and reproduces the core rotation rates of red giants (thick solid blue line in Fig.~\ref{fig:omega_logg}). We implemented the general TS formalism of \cite{eggenberger22} in STAREVOL, using different versions that vary the $C_T$ value (see Sect.~\ref{ref:res_calib}). 
The dynamo operates only when the shear parameter $q$ exceeds a critical threshold value $q_{\rm min}$ required for the magnetic instability to develop. This minimum threshold can be expressed as \citep[Eq.~12 in][]{eggenberger22}:
\begin{equation}
q_{\rm min, T} = C_T^{-1} \left( \frac{N_{\rm eff}}{\Omega} \right)^{(n+2)/2} \left( \frac{\eta}{r^2 \Omega} \right)^{n/4}.
\end{equation}

\subsection{Convective overshooting}
\label{sect:os}

The prescription of \cite{goriely18c} was used for convective overshooting. The overshoot diffusion coefficient, $D_{\rm over}$, follows the expression
\begin{equation}
D_{\rm over} (z) = D_{\rm min} \, \times \, \left( \frac{D_{\rm cb}}{D_{\rm min}} \right)^{(1-z/z^{*})^{p}}
\label{eq:os18},
\end{equation}
where $z^{*} = f_{\rm over} \, H_p \, \ln(D_{\rm cb}) / 2$ is the distance over which mixing occurs, $D_{\rm min}$ is the value of the diffusion coefficient at the boundary $z=z^{*}$, and $p$ and $f_{\rm over}$ are free parameters. 
Below $D_{\rm min}$, it was assumed that $D_{\rm over} = 0$. 
\cite{choplin24} have extensively studied the impact of these overshoot parameters on the development of PIE and i-process nucleosynthesis in AGB stars. 
It was shown that convective overshooting at the top of the thermal pulse can trigger PIEs in higher-mass, higher-metallicity AGB stars. 
Here we considered convective overshooting at the top of the thermal pulse only and adopted $f_{\rm over}=0.04$, $D_{\rm min}=1$~cm$^2$\,s$^{-1}$ and $p=1$ in all models \citep[these are the standard values used in][]{choplin24}.

\begin{figure}[t]
\centering
\includegraphics[width=1\columnwidth]{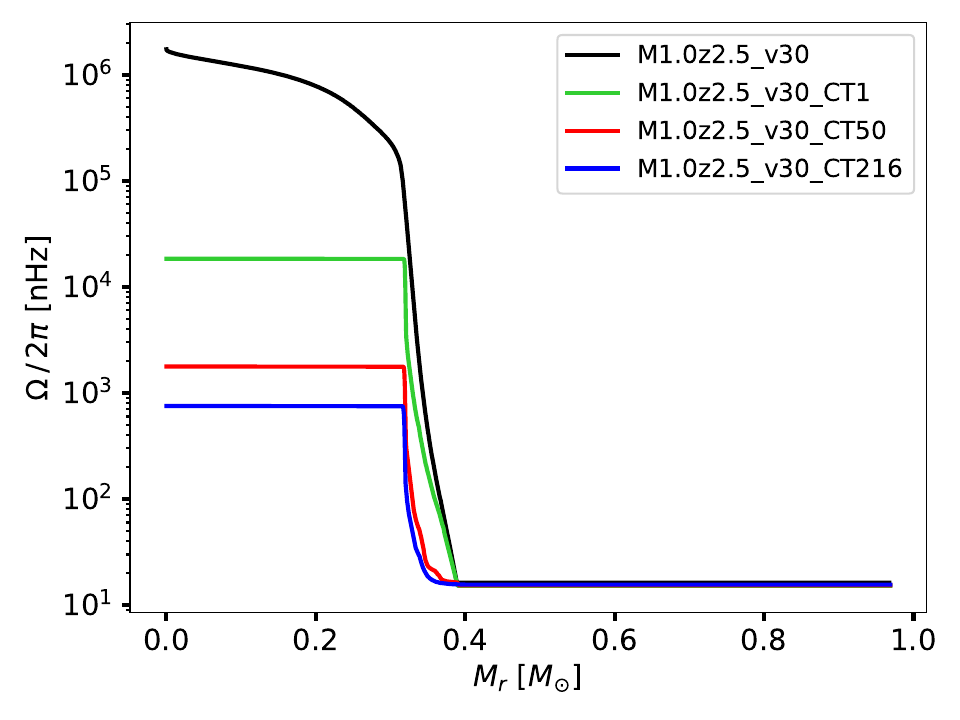}
\caption{Angular velocity profiles for rotating 1.0~\Msun, [Fe/H]~$=-2.5$ models for various values of $C_T$. The structure corresponds to the moment when the convective envelope reaches its deepest extent during the first dredge-up. 
}
\label{fig:omega1}
\end{figure}

\begin{figure*}[t]
\centering
\includegraphics[width=0.66\columnwidth]{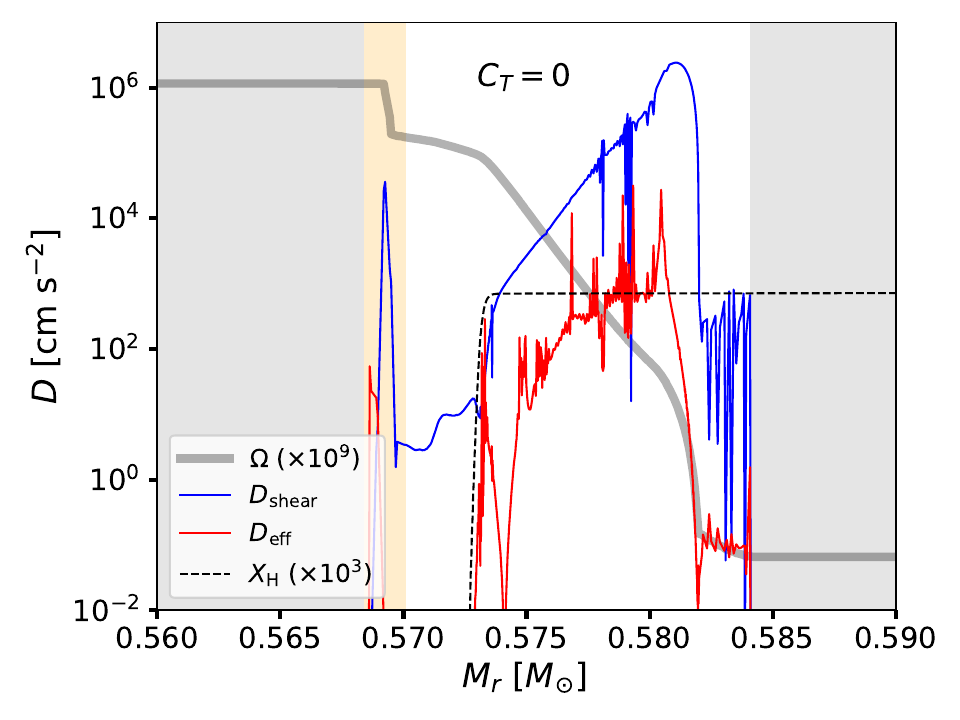}
\includegraphics[width=0.66\columnwidth]{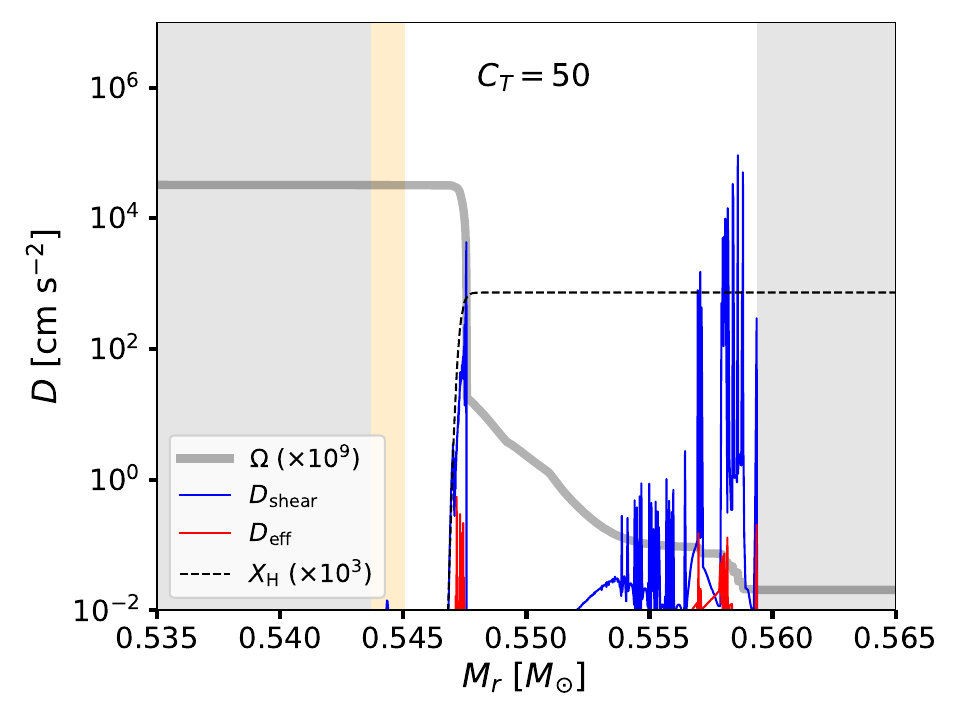}
\includegraphics[width=0.66\columnwidth]{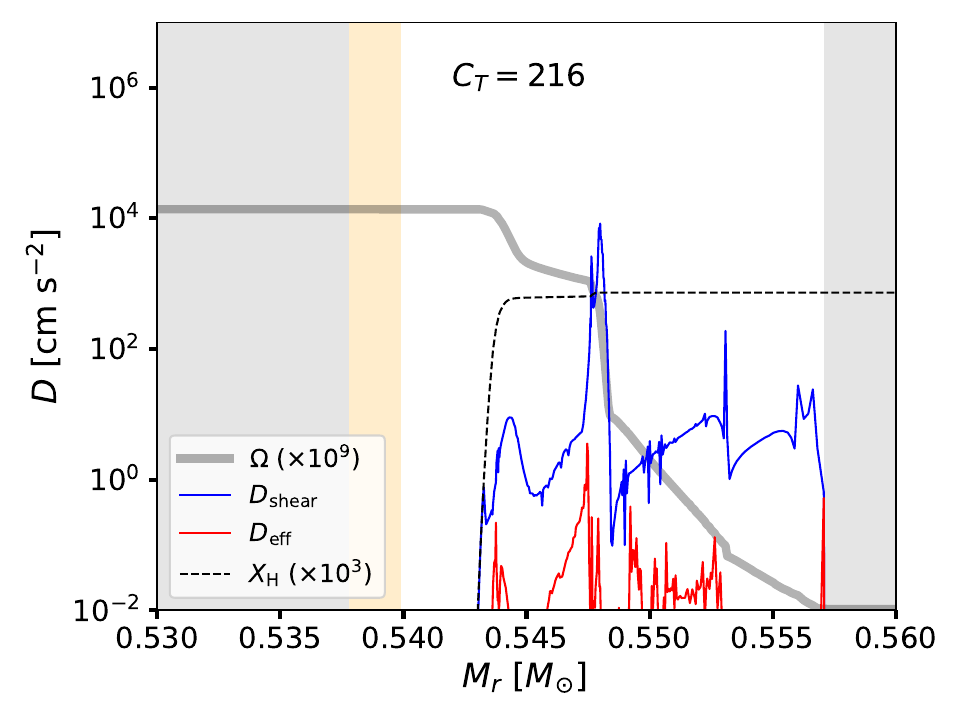}
\caption{
Diffusion coefficients $D_{\rm shear}$ and $D_{\rm eff}$ between the convective thermal pulse and envelope, just prior to the PIE for 1~\Msun, [Fe/H]~$=-2.5$, $v_{\rm ini} = 30$~km~s$^{-1}$ models with $C_T = 0$ (no TS dynamo), $C_T = 50$, and $C_T = 216$. The profiles of angular velocity ($\Omega$) and hydrogen mass fraction ($X_H$, scaled by $10^9$ and $10^3$, respectively) are also shown. Grey- and orange-shaded areas indicate convective and overshoot zones, respectively.
}
\label{fig:dcoeff}
\end{figure*}

\begin{figure*}[t]
\centering
\includegraphics[width=1\columnwidth]{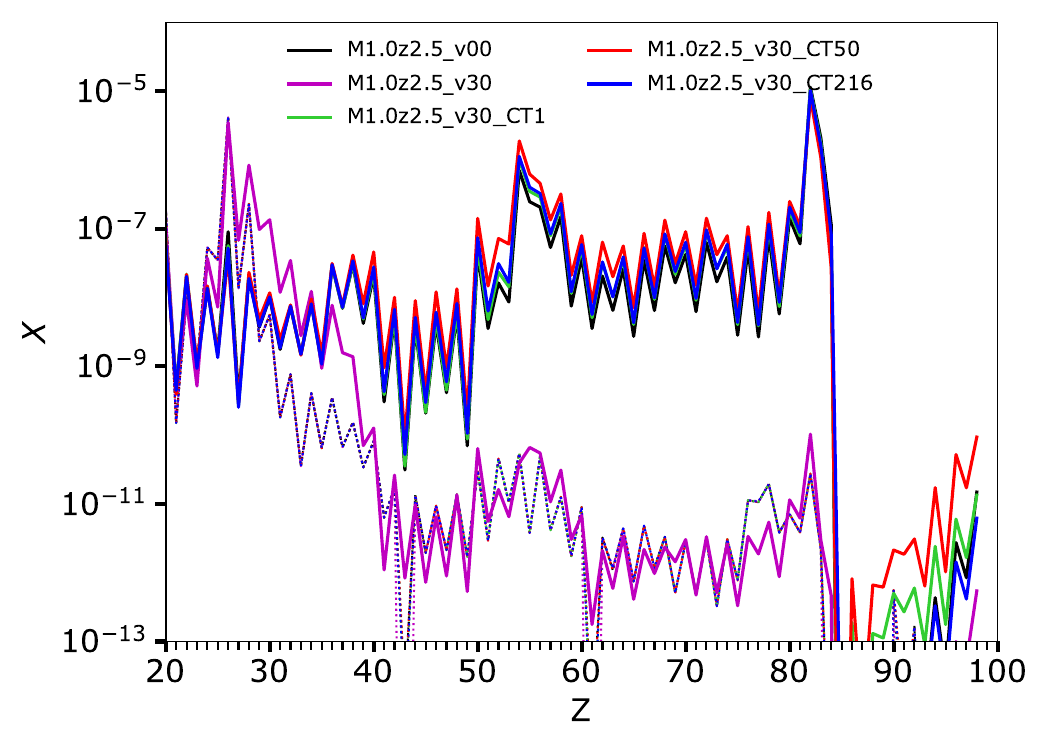}
\includegraphics[width=1\columnwidth]{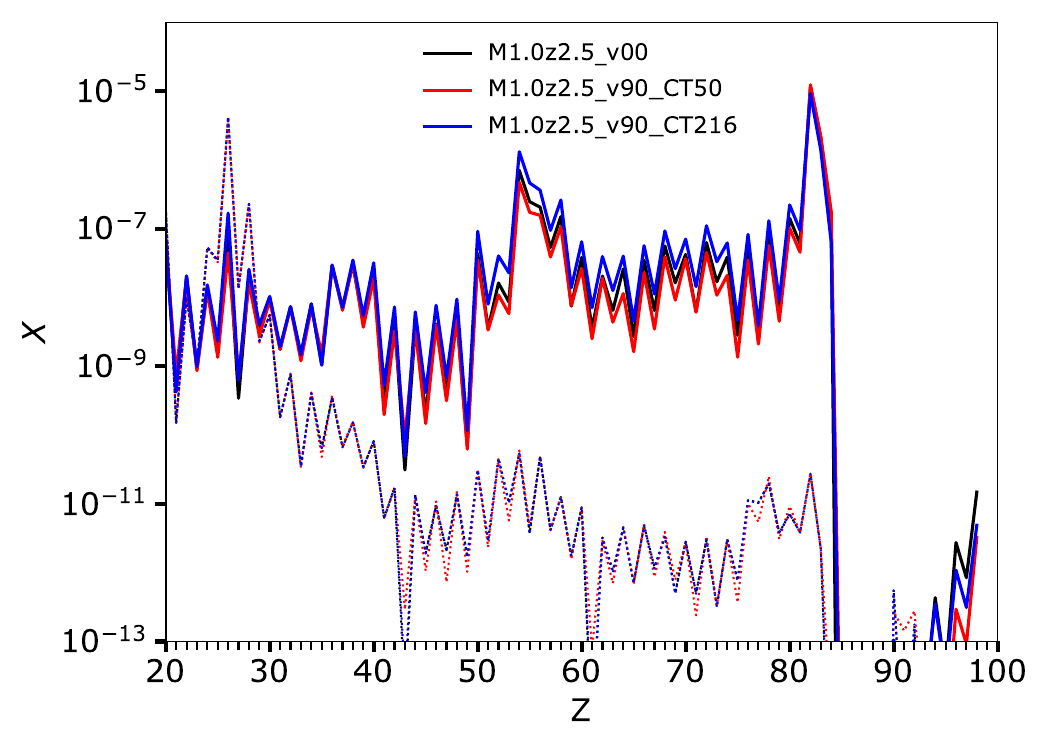}
\caption{
Mass fractions of heavy elements in the convective thermal pulse, just before (dotted lines) and after (solid lines) the main proton ingestion for our 1~\Msun, [Fe/H]~$=-2.5$ models. \textit{Left panel}: Non-rotating model (black) and models with $v_{\rm ini} = 30$~km~s$^{-1}$ and $C_T = 0$ (magenta), 1 (green), 50 (red), and 216 (blue). \textit{Right panel}: Non-rotating model (black) and models with $v_{\rm ini} = 90$~km~s$^{-1}$ and $C_T = 50$ (red) and 216 (blue).
}
\label{fig:mfCT}
\end{figure*}

\begin{figure*}[t]
\centering
\includegraphics[width=2\columnwidth]{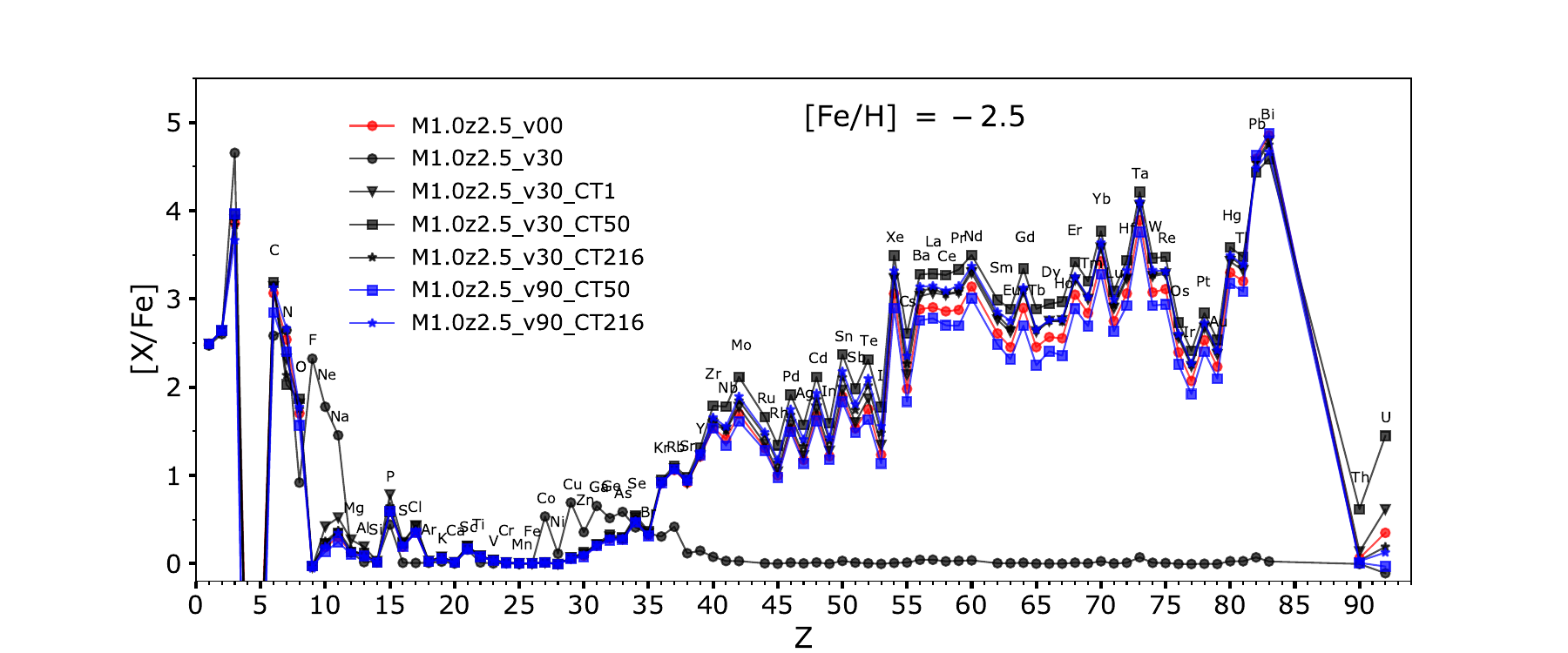}
\includegraphics[width=2\columnwidth]{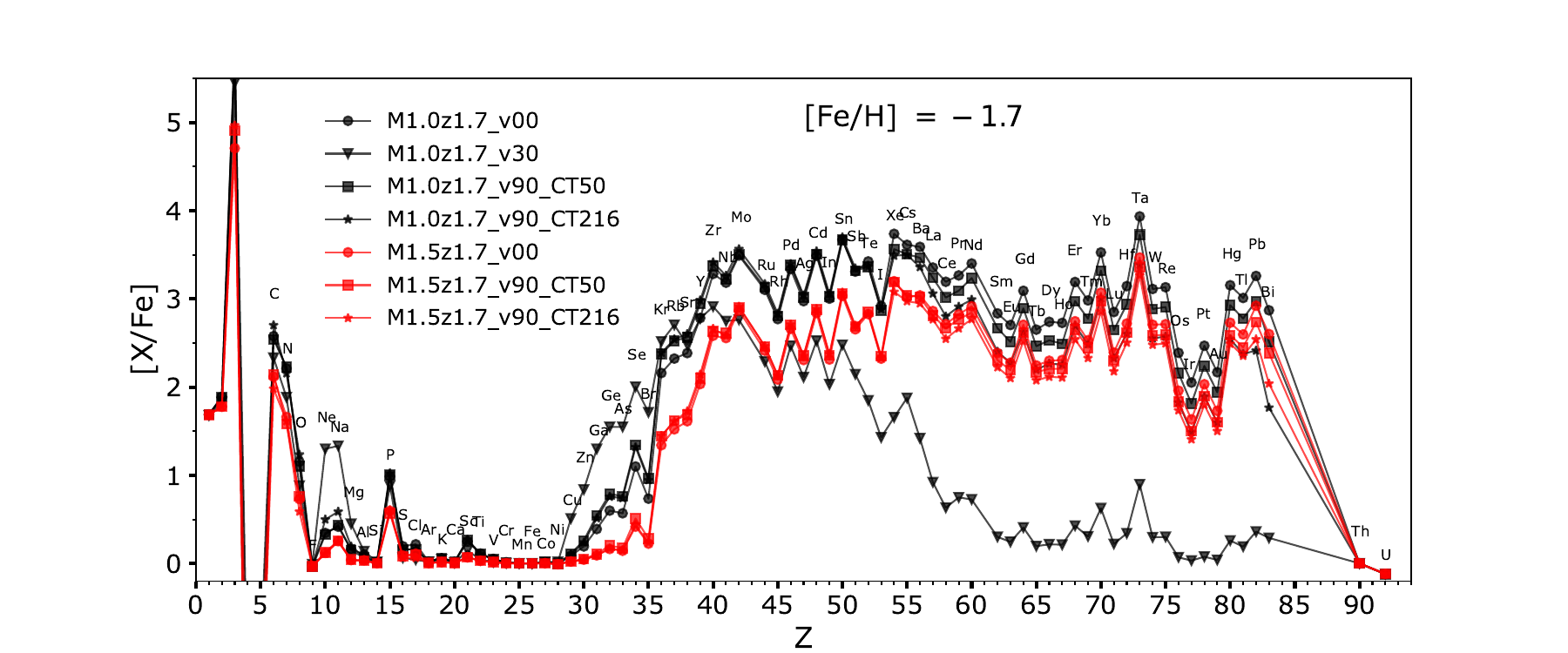}
\caption{Final surface [X/Fe] ratios after the PIE for our [Fe/H]~$=-2.5$ (top) and [Fe/H]~$=-1.7$ (bottom) models. }
\label{fig:xfeROT}
\end{figure*}

\begin{figure*}[t]
\centering
\includegraphics[width=1\columnwidth]{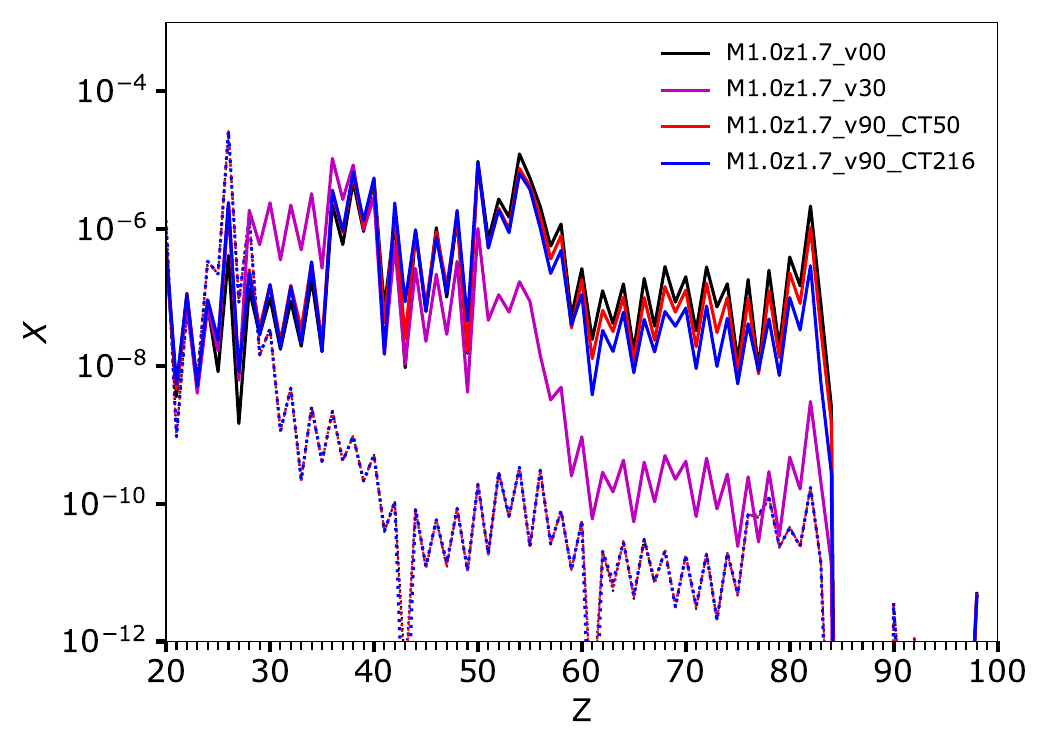}
\includegraphics[width=1\columnwidth]{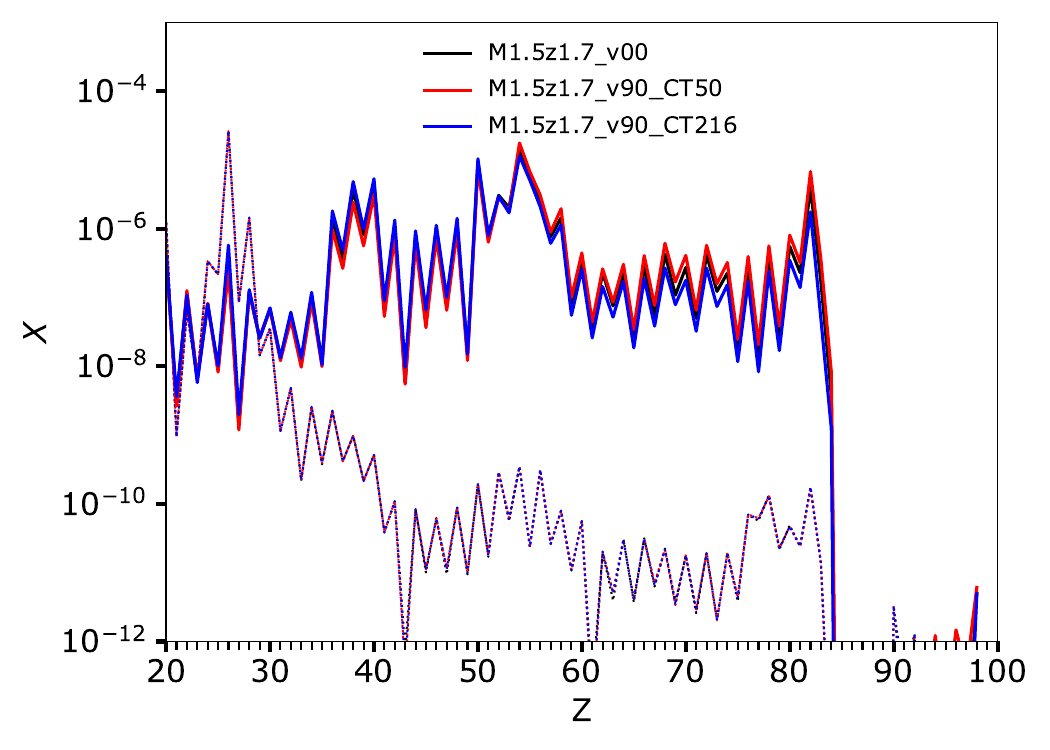}
\caption{Same as Fig.~\ref{fig:mfCT}, but for the [Fe/H]~$=-1.7$ models with $M_{\rm ini} = 1.0$~\Msun\, (left panel) and 1.5~\Msun\, (right panel). Models without rotation (black), with $v_{\rm ini} = 30$~km~s$^{-1}$ (magenta), with $v_{\rm ini} = 90$~km~s$^{-1}$ and $C_T = 50$ (red) and with $v_{\rm ini} = 90$~km~s$^{-1}$ and $C_T = 216$ (blue) are shown.}
\label{fig:mfROT2}
\end{figure*}

\section{Asteroseismic-calibrated version of the Tayler-Spruit dynamo}
\label{ref:res_calib}

To calibrate the modified TS dynamo of \cite{eggenberger22}, we computed 1.1~\Msun\ solar-metallicity models ($Z=0.0134$) with STAREVOL, adopting $v_{\rm ini}=5$~km~s$^{-1}$ \citep[as in][]{eggenberger22} and testing $C_{\rm T}=0$ (no AM transport by magnetic fields), $1$ (original TS dynamo), $50$, and $216$. The surface rotation of all models (dashed black line in Fig.~\ref{fig:omega_logg}) agrees with the observed sub-giant velocities of \citet[][filled black symbols]{deheuvels14} and follows a similar evolution to the \cite{eggenberger22} reference model, computed with the GENEC stellar evolution code (dashed blue line). 

The core rotation rates of sub-giants and giants inferred from asteroseismology are reproduced only by our magnetic models with $C_{\rm T}=50$ (red line) and, to a lesser extent, $C_{\rm T}=216$ (thin solid blue line). From $\log g \simeq 3.75$, our $C_{\rm T}=216$ model rotates about twice as slowly as the \cite{eggenberger22} model (thick solid blue line), whereas the $C_{\rm T}=50$ case recovers their reference track. These discrepancies likely stem from differences in the input physics between GENEC and STAREVOL (e.g. advective vs diffusive treatment of AM transport), whose detailed origin is beyond the scope of this work. Overall, this calibration step confirms that STAREVOL models can reproduce the observed core rotation rates of giants when adopting the generalized TS formalism.

In the following, we examine low-metallicity 1 and 1.5~\Msun\ models (Table~\ref{table:1}) with $C_T=0$ (non-magnetic), 1, 50, and 216. 
Current asteroseismic constraints do not indicate any significant dependence of $C_T$ on stellar mass \citep[e.g. Fig. 3 of][]{eggenberger22}, and the available observations at nearly solar metallicity reveal no clear trend with metallicity. However, since no constraints exist at the low metallicities relevant for this work, exploring different values of $C_T$ remains justified.

\section{Rotational mixing and proton ingestion events}
\label{sect:rotevol}

In magnetic models ($C_T > 0$), core-envelope coupling transfers AM outwards, causing the core rotation rate to steadily decrease rather than rise, as in non-magnetic rotating models. Figure~\ref{fig:omega1} shows the angular velocity profiles of 1.0~\Msun, [Fe/H]~$=-2.5$ models with $v_{\rm ini} = 30$~km~s$^{-1}$ at the beginning of the red giant branch (RGB) for different values of $C_T$. The non-magnetic case (black) exhibits strong core-envelope differential rotation ($\sim 5$ dex), while in magnetic models the coupling strengthens with increasing $C_T$, leading to progressively slower core rotation rates.

At fixed initial mass and metallicity, our models show comparable evolution through the main-sequence, RGB, and AGB phases, with similar thermal pulse properties and envelope masses, independently of the values of $C_T$ and $v_{\rm ini}$. For example, in our seven [Fe/H]~$=-2.5$ models, PIEs occur under nearly identical conditions and lead to similar i-process nucleosynthesis, except in the non-magnetic rotating case (\texttt{M1.0z2.5\_v30}; see Sect.~\ref{sect:weaki}).

Importantly, we find that rotational mixing is too weak in all our rotating models to trigger the occurrence of PIEs on its own. 
As an example, we considered the \texttt{M1.0z1.7\_v90\_CT216} model during the first thermal pulse and switched off overshooting, leading to no PIE occurring\footnote{Without additional mixing processes (e.g. overshoot), PIEs are expected in AGB models with $M_{\rm ini} \lesssim 2.5$~\Msun\ and [Fe/H]~$<-2$ \citep[Fig.~12 in][]{choplin24}.}.
At the pulse maximum extent, the distance between the top of the convective pulse and the first H-rich layer ($X_H > 10^{-3}$) is $r = 0.032~R_\odot$. With a pulse duration of $\tau \sim 400$~yr, the diffusion coefficient needed to connect the H-rich layers to the convective zone is $D \sim r^2 / \tau \sim 4 \times 10^8$~cm$^2$~s$^{-1}$ $-$ several orders of magnitude higher than the typical $D_{\rm shear}$ in magnetic rotating models (Fig.~\ref{fig:dcoeff}, middle and right panels). Non-magnetic models are also far from these values, especially in the zone of interest, i.e. below the H-rich layers (left panel). This demonstrates that rotation-induced mixing during the AGB phase is insufficient to trigger PIEs by itself.

\section{Impact of rotation on i-process nucleosynthesis and Fluorine synthesis}
\label{sect:ipro}

\subsection{The dilution procedure}

To estimate the surface abundances resulting from i-process nucleosynthesis in our 1 \Msun\,, [Fe/H]~$=-2.5$ AGB models, we adopted the dilution procedure described in \citet[][Sect.~3.1]{martinet24}. During a PIE, heavy-element nucleosynthesis occurs prior to the splitting of the convective pulse \citep{choplin22a, choplin24}. Since the post-split evolution (including mixing with the convective envelope) does not substantially alter these abundances, a computationally efficient approach was employed: the mean abundances of each element in the upper part of the convective pulse, just after the split, were diluted into the convective envelope using a fixed dilution factor calibrated on a full non-rotating 1 \Msun\,, [Fe/H]~$=-2.5$ AGB model. 
Because the effect of rotation on the structure is small (Sect.~\ref{sect:rotevol}), this calibration can be safely used in our rotating models -- except for the \texttt{M1.0z2.5\_v30} model, however, which behaves differently during the PIE (Sect.~\ref{sect:weaki}).
This method yields accurate approximations of the final surface abundances without the need to evolve the model through the complete envelope mixing phase. For more details on the implementation and accuracy of this method, we refer to Sect.~3.1 of \cite{martinet24}.\\

\subsection{The [Fe/H]~$=-2.5$ models}

During the PIE, our [Fe/H]~$=-2.5$ models reach peak neutron densities in the range $8.1 \times 10^{13} < N_{\rm n,max} < 8.5 \times 10^{14}$~cm$^{-3}$ (Table~\ref{table:1}). Despite this variation, all models $-$ except \texttt{M1.0z2.5\_v30} (see Sect.~\ref{sect:weaki}) $-$ undergo very similar i-process nucleosynthesis (Fig.~\ref{fig:mfCT}).

The i-process converts mostly Fe $-$ whose mass fraction drops from $4 \times 10^{-6}$ before the PIE to $\sim 10^{-7}$ after in the convective pulse $-$ into heavier elements. Most trans-iron elements increase by $3-4$~dex (Fig.~\ref{fig:mfCT}), with significant production of Sr ($Z=38$), Zr ($Z=40$), Sn ($Z=50$), Xe ($Z=54$), Cs ($Z=55$), Ba ($Z=56$), and Pb ($Z=82$). The processed material is subsequently diluted into the convective envelope, enriching the AGB surface; the final [X/Fe] ratios are shown in Fig.~\ref{fig:xfeROT} (top panel). In most models, actinide production (especially Th and U) remains low because convective overshooting, included in all our models, was shown to slightly weaken the i-process, limiting the synthesis of actinides \citep{choplin24}. An exception is the \texttt{M1.0z2.5\_v30\_CT50} model, which reaches the highest neutron density ($8.5 \times 10^{14}$~cm$^{-3}$) and produces [Th/Fe]~$=0.61$ and [U/Fe]~$=1.44$ (Fig.~\ref{fig:xfeROT}, top panel).

\subsubsection{The poisoning effect of primary $^{22}$Ne on i-process nucleosynthesis}
\label{sect:weaki}

Although its peak neutron density is comparable to that of other [Fe/H]~$=-2.5$ models (Table~\ref{table:1}), the \texttt{M1.0z2.5\_v30} model experiences only a very weak i-process during the PIE, which occurs during the very first TP. As a result, it exclusively produces elements with $Z<40$, resembling a convective s-process nucleosynthesis (magenta pattern in the left panel of Fig.~\ref{fig:mfCT}). The reason for this is discussed below.

After the main-sequence and RGB phases, our models undergo helium flash, which ignites off-centre at $M_r \simeq 0.2$~\Msun\ due to neutrino losses. A large convective zone ($M_r \simeq 0.2-0.5$~\Msun) forms, producing significant amounts of $^{12}$C from $^{4}$He. Once this convective zone recedes, He burning continues in the core, leaving a $^{12}$C-rich radiative layer beneath the H-burning shell. In the \texttt{M1.0z2.5\_v30} model, rotational mixing mixes $^{12}$C into the H-burning shell, boosting the CNO cycle and generating primary\footnote{Synthesized from the initial H and He contents, as opposed to secondary production from the initial metal (elements heavier than He) content.} $^{14}$N and $^{13}$C.
During the subsequent evolution, when the He-burning shell moves through $^{14}$N-rich layers, $^{22}$Ne is massively produced by $^{14}$N($\alpha,\gamma$)$^{18}$F($\beta^+$)$^{18}$O($\alpha,\gamma$)$^{22}$Ne. Just before the first thermal pulse (where a PIE occurs), the $^{22}$Ne mass fraction peaks at $1.7\times10^{-2}$ in \texttt{M1.0z2.5\_v30}, compared with $\lesssim 5\times10^{-5}$ in other [Fe/H]~$=-2.5$ models. During the PIE, $^{22}$Ne($n,\gamma$) reaction has among the strongest fluxes and acts as a severe poison. It results in a weak i-process that synthesizes mainly elements with $Z<40$ (Figs.~\ref{fig:mfCT}, \ref{fig:xfeROT}). Consequently, the final surface Ne abundance is high in this model (Fig.~\ref{fig:xfeROT}, top panel). 

The $^{22}\mathrm{Ne}(n,\gamma)$ reaction competes with $^{22}\mathrm{Ne}(p,\gamma)$, for which we adopted the rate of \cite{longland10}. Recently, \cite{lennarz20} derived a new experimental rate for the latter reaction, which is higher by a factor of $\sim 3$ compared to \cite{longland10} at He-burning temperatures. As a test, we  recomputed the \texttt{M1.0z2.5\_v30} model during the PIE using the $^{22}\mathrm{Ne}(p,\gamma)$ rate increased by a factor of three. We find that such an enhancement leads to an overproduction by a factor of $\sim 2-3$ of elements with $30 < Z < 40$ in the thermal pulse after the PIE.

In the non-rotating [Fe/H]~$=-2.5$ model, no additional $^{22}$Ne is produced because mixing is absent. In magnetic rotating models, the TS dynamo flattens the angular velocity profile (Fig.~\ref{fig:omega1}), which greatly suppresses shear mixing—the main driver of chemical transport in non-magnetic rotating models—since it depends on the gradient of $\Omega$ (Eq.~\ref{eq:dshm97}). Consequently, final surface Ne ($Z=10$) abundances remain much lower in both non-rotating and magnetic models (Fig.~\ref{fig:xfeROT}, top panel).

\subsubsection{Fluorine and Sodium synthesis during the PIE of the \texttt{M1.0z2.5\_v30} model}
\label{sec:fluorine}

The \texttt{M1.0z2.5\_v30} model undergoes a weak i-process (Sect.~\ref{sect:weaki}) but is the only model that produces significant Fluorine ([F/Fe]~$\simeq 2.5$, Fig.~\ref{fig:xfeROT}). 
The $^{19}$F isotope originates from primary $^{14}$N synthesized prior to the AGB phase (Sect.~\ref{sect:weaki}) through four roughly equally important reaction chains:
\begin{itemize}
\item $^{14}$N($\alpha$,$\gamma$)$^{18}$F($n$,$\alpha$)$^{15}$N($\alpha$,$\gamma$)$^{19}$F,  
\item $^{14}$N($\alpha$,$\gamma$)$^{18}$F($\beta^+$)$^{18}$O($p$,$\alpha$)$^{15}$N($\alpha$,$\gamma$)$^{19}$F,
\item $^{14}$N($\alpha$,$\gamma$)$^{18}$F($n,p$)$^{18}$O($p,\alpha$)$^{15}$N($\alpha$,$\gamma$)$^{19}$F,
\item $^{14}$N($\alpha$,$\gamma$)$^{18}$F($\beta^+$)$^{18}$O($n$,$\gamma$)$^{19}$O($\beta^-$)$^{19}$F.
\end{itemize}

Because of the high neutron density, the reaction $^{19}\mathrm{F}(n,\gamma)^{20}\mathrm{F}$ is also efficient, causing partial destruction of $^{19}\mathrm{F}$. However, the net outcome remains a substantial production of $^{19}\mathrm{F}$.

Sodium is also efficiently produced in this model ([Na/Fe]~$\simeq 1.5$, Fig.~\ref{fig:xfeROT}) through the reaction chain $^{22}$Ne($n,\gamma$)$^{23}$Ne($\beta^-$)$^{23}$Na, with most of the $^{22}$Ne originating from the primary $^{14}$N synthesized earlier (Sect.~\ref{sect:weaki}).
Thus, substantial F, Ne, and Na production during PIEs requires prior synthesis of primary $^{14}$N. This occurs only in our rotating, non-magnetic model, which is disfavoured by asteroseismic constraints (Fig.~\ref{fig:omega_logg}), even though no asteroseismic observations exist at such low metallicities.

\subsection{The [Fe/H]~$=-1.7$ models}

As for the [Fe/H]~$=-2.5$ models, including rotation and magnetic fields in the [Fe/H]~$=-1.7$ models results in evolution and nucleosynthesis (Fig.~\ref{fig:xfeROT} and \ref{fig:mfROT2}) nearly identical to the non-rotating cases. 
Overall surface enrichment is lower in the 1.5~\Msun\ models than in the 1~\Msun\ models, because the PIE products are diluted into a more massive convective envelope ($\sim 0.2$ vs $\sim 0.7$~\Msun, respectively).

The rotating non-magnetic \texttt{M1.0z1.7\_v30} model undergoes a weaker i-process nucleosynthesis (bottom panel of Fig.~\ref{fig:xfeROT},  and left panel of Fig.~\ref{fig:mfROT2}) for the same reasons as the \texttt{M1.0z2.5\_v30} model (Sect.~\ref{sect:weaki}). 
However, compared to the  \texttt{M1.0z2.5\_v30} model (Fig.~\ref{fig:xfeROT}), a larger amount of heavy elements is produced because the PIE develops during the third rather than the first thermal pulse, as in the lower metallicity model. Although the $^{22}$Ne abundance during the PIE is similar in both cases, most of the primary $^{14}$N is consumed in the two preceding TPs, thereby reducing its poisoning effect and allowing for more efficient synthesis of heavy elements.

Another difference is that, unlike the \texttt{M1.0z2.5\_v30} model, no fluorine enhancement is noticed at the surface of the \texttt{M1.0z1.7\_v30} model (Fig.~\ref{fig:xfeROT}, bottom panel).
Owing to the large amount of primary $^{14}\mathrm{N}$ synthesized prior to the AGB phase, $^{19}\mathrm{F}$ is efficiently produced during the first two TPs (where no PIE occurs) through the reaction chain $^{14}\mathrm{N}(\alpha,\gamma)^{18}\mathrm{F}(\beta^+)^{18}\mathrm{O}(p,\alpha)^{15}\mathrm{N}(\alpha,\gamma)^{19}\mathrm{F}$ with protons supplied by  $^{14}\mathrm{N}(n,p)^{14}\mathrm{C}$. At this stage, the neutron flux is too low to significantly destroy $^{19}\mathrm{F}$ via the $^{19}\mathrm{F}(n,\gamma)^{20}\mathrm{F}$ channel. But the $^{19}\mathrm{F}$ synthesized during these first two TPs does not reach the stellar surface because the third dredge-ups are not deep enough. 
During the subsequent third TP, when the PIE takes place, the production of $^{19}\mathrm{F}$ becomes inefficient because only a small amount of $^{14}\mathrm{N}$ remains. Instead, the destruction channel $^{19}\mathrm{F}(n,\gamma)^{20}\mathrm{F}$ strongly dominates, ultimately leading to the nearly complete depletion of $^{19}\mathrm{F}$. 
Hence, efficient fluorine production requires not only prior synthesis of primary $^{14}$N but also that the PIE occurs during the very first TP.

 \section{Conclusions}
\label{sect:concl}

In this study, we explored the effect of rotation on i-process nucleosynthesis in AGB stars. To incorporate constraints from asteroseismic observations, we implemented in STAREVOL a modified version of the TS dynamo, as recently proposed by \citet{eggenberger22}, designed to reconcile rotating models with observational data available for red giant stars. We computed both classical rotating models (without the dynamo) and magnetic rotating models (with the dynamo).

We found that i-process nucleosynthesis is strongly suppressed in rotating AGB models that employ the classical treatment of AM transport based purely on hydrodynamical processes and no magnetic fields \citep{zahn92}.
This is due to the production of primary $^{14}$N prior to the AGB phase, which is subsequently converted into $^{22}$Ne that acts as a strong neutron poison during the PIE via the $^{22}$Ne($n,\gamma$) reaction. As a result, only elements with $Z<40$ are synthesized, and their enhancement is modest ([X/Fe]~$< 1$~dex). In addition, rotational mixing alone can lead to significant production of fluorine and sodium during the PIE through four reaction chains originating from $^{14}$N. However, efficient fluorine synthesis occurs only if the PIE develops during the very first thermal pulse.

In contrast, in rotating magnetic models, strong core-envelope coupling suppresses shear mixing, preventing production of primary $^{14}$N and the associated excess of $^{22}$Ne. Consequently, i-process nucleosynthesis in these models closely mirrors those of non-rotating stars. This behaviour is largely independent of the rotation rate or the strength of the TS dynamo (parametrized by $C_T$) and is consistent across different metallicities ([Fe/H]~$= -2.5$ and $-1.7$) and initial masses (1 and 1.5 \Msun ). 
Although overshooting is crucial for initiating PIEs \citep[especially at higher masses and metallicities][]{choplin24}, we find that rotational mixing during the AGB phase is too weak to affect their occurrence.

Internal magnetic fields generated by the TS dynamo can help reconcile models with asteroseismic observations, but they cannot fully explain the missing AM transport in stars by adopting a single value for the calibration parameter. The internal rotation profiles of, for example, $\gamma$ Doradus stars \citep{moyano24} and white dwarfs \citep{denhartogh20} have indeed been shown to be consistent with magnetic models that use a lower value of the TS dynamo calibration parameter than that needed to reproduce the core rotation rates of red giants. Additional processes $-$ such as internal gravity waves \citep{pincon17}, mixed modes \citep{belkacem15,bordadagua25}, or azimuthal magneto-rotational instability \citep{moyano23b, meduri24} $-$ could also play a role in the internal transport of AM. These mechanisms could also affect AGB nucleosynthesis and should be investigated in future studies.

\begin{acknowledgements}
This work was supported by the Fonds de la Recherche Scientifique-FNRS under Grant No. IISN 4.4502.19. L.S. and S.G. are senior F.R.S-FNRS research associates. A.C. is post-doctorate F.R.S-FNRS fellow. P.E. acknowledges support from the SNF grant No 219745 (Asteroseismology of transport processes for the evolution of stars and planets).
\end{acknowledgements}

\bibliographystyle{aa}
\bibliography{astro.bib}

\end{document}